# Large photoresponse of Cu:TCNQ nanowire arrays formed as aligned nanobridges.

Rabaya Basori†*, K. Das†, Prashant Kumar‡, K.S.Narayan‡, A. K. Raychaudhuri†

†*Theme Unit of Excellence on Nano Device Technology, S.N. Bose National Centre for Basic Sciences, Salt Lake, Kolkata-700098, India*

‡*Jawaharlal Nehru Center for Advanced Scientific Research, Jakkur Bangalore-560 064, India*

E-mail: (rabaya@bose.res.in, arup@bose.res.in)

We report for the first time a large photoresponse in an array of charge transfer complex Cu:TCNQ nanowires (average diameter 30 nm) fabricated as a nanobridge device. The device shows highest photoresponse for excitation with 405 nm light which matches with its absorption peak. The current gain at zero bias can reach ~$10^4$ with an illumination power density of $2 \times 10^6$ W/m$^2$. The zero bias responsivity is ~0.3 mA/W which increases on applying bias reaching 1.0 A/W or more for a bias of 2.0 Volt. Dark and illuminated I-V data are analyzed by two back-to-back Schottky diodes model, which shows the predominant photocurrent in the device arising from the photoconductive response of the nanowires.

In recent years, a variety of semiconductor nanowires (NWs) have been synthesized and used as basic building blocks for the development of electronic and optoelectronic nanodevices[1-6]. In particular there are exciting reports of optoelectronic nanodevices made from single nanowires or arrays of nanowires.[4,7] In this paper we report fabrication and characterization of an optoelectronic device that is based on well-aligned array of nanowires of the charge transfer complex Cu:TCNQ. The nanowires with diameter $\approx$ 30nm were laterally grown by vapour deposition on prefabricated electrodes allowing formation of nanobridges with electrode separation $\approx$ 1.5 μm. The arrays so formed show a large photo responsivity $\Re$ (defined as Current generated/power absorbed) of nearly 1.0 A/W at a bias of 2V and even a zero bias a finite responsivity $\Re(V=0)$ of 0.3mA/W could be reached in arrays with about 40-80 nanowires. The spectral response has a maximum around $\lambda \approx 405 nm$. The reported photo-response is large enough that Cu:TCNQ nanowires can be considered for making visible photo-detectors.

The charge transfer complex Cu:TCNQ has attracted considerable recent attention for its electrical resistive state switching properties that show bias driven transition with memory from a high resistance state to a low resistance state. This makes the material attractive for making MEMISTOR and non-volatile RAM.[8] The present report of large photo-response in CuTCNQ NW arrays, thus adds an opto-electronic element to the existing electronic switching property and raises the possibility of opto-electronic control to a MEMISTOR device. The phenomena represented in this paper are mainly controlled by the generation of charge carrier due to illumination of laser light that leads to substantial photoconductivity.

There are past reports of resistive state switching in CuTCNQ films using high power laser illumination and its use as an enabling method for optical writing on this material.[9,10] The reported works, mentioned above have been done on films and there are no reports of photoresponse investigations in Cu:TCNQ nanowire arrays as has been done here.

The fabrication route followed, namely growth of nanowires within prefabricated electrodes /electrode arrays, has the possibility of fabrication of high density arrays and it allows direct integration of growth of nanowires with the device fabrication process[5,11]. In this process, the main device architecture and electrode can be fabricated before the growth of nanowires. As a result there will be a little chance to nanomaterial deterioration due to post treatment. The nanobridge type NW arrays fabricated here allow direct use as photo-conductive nanodevice without any additional step.

A schematic diagram and fabrication process of the Cu:TCNQ nanobridge device on Si/SiO$_2$ substrate are illustrated in Figure 1a. More details of the fabrication process are given in supplementary (Figure S1). The electrode patterns were formed on e-beam written patterns on PMMA resist followed by Cu and Au evaporation sequentially and then lift-off. TCNQ was evaporated in vacuum on the electrode patterned substrates maintained at around 130$^0$C temperature. Cu:TCNQ NWs grow laterally from the sidewall of Cu electrodes that are exposed. SEM image of a nanobridge device with horizontal arrays of the nanowires is shown in figure 1b. The average diameter of the nanowires that bride the two electrodes is around 30nm. The quality of the nanowires formed has been been tested by a number of electron microscopy, X-Ray diffraction and Raman Spectroscopy. The results are given in Supplemantary.

The zero-bias photocurrent was measured with a 405 nm diode Laser as a Source coupled to a microscope using a Lock-In Amplifier and a mechanical chopper and. The illumination power was varied by changing neutral density filters. The spectral response over a wide range of wavelength (in visible) was measured using a halogen lamp and a monochromator. The measurement system was calibrated using a power-meter. The electrical measurements (including a temperature dependent resistivity measurement) were made using Source Meter (Keithley 2400) and Pico-Ammeter (Keithley 6485). All the photocurrent measurements have been done at room temperature and in atmospheric pressure. The resistivity of the NW in such a device was measured separately in a cryostat as a function of temperature (T) and is shown in figure 1c.

Figure 2 shows a typical *I-V* data taken on such a nanobridge device at dark and also under continuous illumination (power density $P_{opt}$ = 6.6x10$^6$W/m$^2$) of wavelength λ= 405nm. The data shown were taken on an array with 80 nanowires. The *I-V* data of such a device is not symmetric which arises due to asymmetric contacts which we quantify below. The asymmetry in contacts occurs because wires were made to grow preferentially from one electrode of the bridge and terminate on the other. The electrode, from which growth starts, was selected by choosing the angle of orientation of the substrate during growth. The illumination at an applied bias of 2V can have an enhancement of the current in the device by nearly 2 orders. The data reported in this work were taken at an illumination of 405nm because there response shows a peak at this wavelength. Separate studies on an ensemble of nanowires show that the absorption reaches a peak around this wavelength. This links the photo-response to fundamental carrier generation in the material.

The dark as well as illuminated *I-V* data have been analyzed considering the device as metal-insulator-metal (MSM) structure where Cu electrodes act as metal electrodes and Cu:TCNQ NWs as semiconductor. The current through the device has two components, zero bias photo- current $I_{Ph}(V=0)$ and a bias dependent current *I(V)*. The zero bias photo current is measured separately. The bias dependent current *I(V)* through the MSM device is fitted with a model of two back-to-back Schottky diodes connected by a series resistance R, representing the resistance of the NWs between two electrodes. Equation used for fit is:[12]

$$I(V) = I_0 \exp\left(\frac{qV'}{\eta kT} - 1\right) \frac{\exp\left(\frac{-q(\phi_1 + \phi_2)}{kT}\right)}{\exp\left(\frac{-q\phi_2}{kT}\right) + \exp\left(\frac{-q\phi_1}{kT}\right)\exp\left(\frac{qV'}{\eta kT}\right)} \quad (1)$$

where, $V' = V - IR$, $R$ being the series resistance, $\phi_1$ and $\phi_2$ are the barrier heights associated with two contacts (M's) and $\eta$ is the ideality factor. In the equation above $\phi_1$ refers to terminal with V +ve. This is the electrode from which growth starts (marked in Figure 1b). $I_0$ arises from thermionic emission and does not change on illumination. The bias dependent $I(V)$ is added to experimentally observe zero bias photocurrent $I_{Ph}(V = 0)$ to obtain the total device current $I$. If the two M contacts in the MSM structure are identical ($\phi_1 = \phi_2$) and the $I - V$ curves will be symmetric. However, unequal barrier heights ($\phi_1 \neq \phi_2$) at the two contacts will lead to asymmetric $I - V$ curves as observed. A simple schematic of the barriers and the MSM device model used in the analysis is given in the inset of figure 2. From the fit of the experimental $I - V$ data using equation 1 (shown in figure 2), we obtain $\phi_1 \approx 1.3 \text{meV}$ and $\phi_2 \approx 0.2 \text{eV}$ in the dark with $R = 8$ k$\Omega$. This shows very small barrier height in one end of the nanowire electrode junction where growth starts indicating that the contact is almost ohmic. At other end where NWs make contact to the electrode and the barrier height is larger. The value of the series resistance $R$ in dark obtained from the fit, matches well with that calculated from the resistivity in dark that has been measured separately (see figure 1c)

On illumination at 405nm, the barrier heights change marginally. While $\phi_1$ remains nearly unchanged (1.3 *meV* to 1.0 *meV*) and there is a small lowering of $\phi_2$ from 0.20 *eV* to 0.18 *eV*. The main change on illumination occurs for the series resistance that shows a significant reduction. The value of the series resistance under illumination R =0.8 kΩ (obtained from the fit) is an order lower than the dark value. The suppression of the series resistance *R* is due to the photoconductivity in the NWs between the two contacts. In this context it is noted that in many nanowire based photo-detectors (working as MSM device) the response under illumination can arise from the photoconductivity of the NW and also from significant lowering of the barrier under illumination.[13-14] In the nanowire arrays grown directly on Cu electrodes, we find that the barrier being low the contribution arising from lowering of barrier is also low and most of the response arises from the photoconductivity of the wire.

Figure 3 represents the photo-response of the array of NWs at zero bias, taken at a wavelength λ= 405nm at different incident powers using a mechanical chopper (frequency 130 Hz.). The photo-response time shown in figure 3 is limited by the speed of the mechanical chopper. At zero bias photocurrent under illumination is $10^4$ times more than the dark current. There is no persistent photo-current as the dark current is fully recovered when the illumination is turned off. The zero bias photo-current, $I_{ph}$ ($V = 0$) increases with the incident optical power density $P_{opt}$ as shown in the inset of figure 4a. It was found to have dependence of photocurrent on optical power as:

$$I_{Ph}(V = 0) \propto P_{opt}^{\gamma} \quad (2)$$

From the fit we find $\gamma \approx 0.15$. The exponent $\gamma \ll 1$ in the fit is a result of the complex process of electron-hole generation, trapping, and recombination within the NW's.[4] The small value of γ of CuTCNQ NWs may indicate existence of abundant trap states in the gap with broad distribution in energy. In case of nanowires, we also have large carrier generation rate in a confined volume which can also increase the non-geminate recombination cross-section and thus reduce $\gamma$.[4]

The responsivity $\Re$ of the nanowire device has been measured. It is defined as $\Re = \frac{I_{ph}}{P_{sample}}$, where $P_{sample}$ is the power absorbed by the nanowires of the array. We assume that all the power falling on the surface of the NW's is absorbed so that the total power $P_{sample} = P_{opt} A N$. N is the number of the NW's in a particular device and A is the average area of a NW. This underestimates the responsivity because not all the power incident on the NW's is absorbed. Due to sub-linear dependence of the photo-current on the incident power, zero bias responsivity $\Re(V=0)$ decreases with increase of illumination power. $\Re(V=0)$ at low power ($P_{opt} = 7.8 \times 10^4$ W/m$^2$) is 2.3 mA/W and it decreases to 0.3 mA/W at $P_{opt} = 2.0 \times 10^6$ W/m$^2$. However, application of bias enhances the responsivity $\Re$ significantly. At V=2 volt, $\Re = 1 A/W$. The variation of responsivity with bias is shown in Figure 4. For bias V > 2 volts, $\Re$ was found to have linear dependence on V and it reaches $\Re = 6 A/W$ at 6V. The device can be operated upto 10V.

To summarize, we have demonstrated that a viable nanobridge type device can be made from Cu:TCNQ NWs that shows very large photoresponse. The nanowires used have diameter ≈ 30nm. The photo-response peaks at 405nm, where the absorption shows a maximum in Cu:TCNQ nanowires. The photo-current under illumination has a zero bias component where an enhancement $\geq 10^4$ over dark current can be seen. The responsivity $\Re$ has a bias dependence which can reach equal to or more than 6 A/W for an applied bias of 6 Volt. The photocurrent primarily occurs due to generation of photocarrier under illumination. This is the first report of photo-response including zero bias response from NWs arrays made of this charge transfer complex material like Cu:TCNQ that is well known for its bias driven resistive state transition. This study raises the possibility of optical control of MEMISTORS and non-volatile RAM's made from this material.

*Acknowledgement:* The work is carried out under the projects Centre for Nanotechnology and Theme Unit of Excellence in nanodevices funded by the Nanomission, Dept. of Science and Technology, Government of India. Prof. A. K. Raychaudhuri and Dr. Kaustuv Das acknowledging project UNANST-II for partial manpower support.

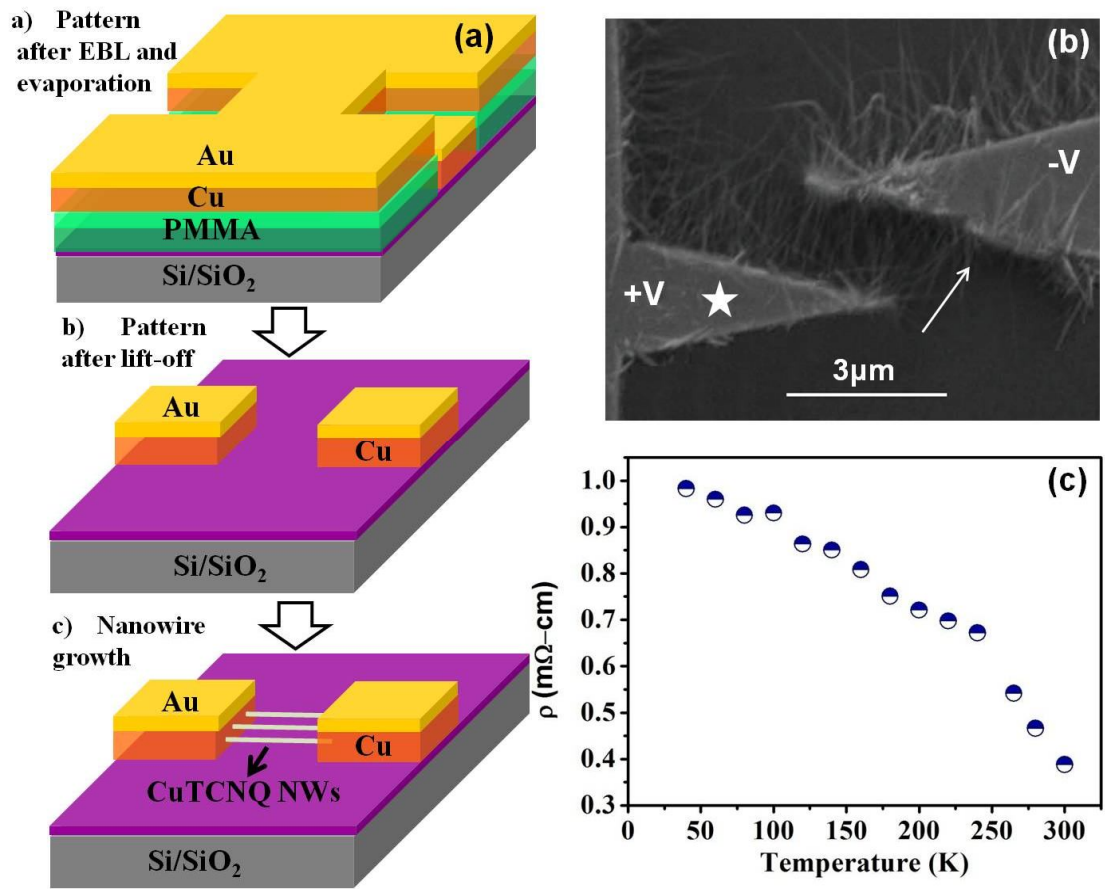

FIG. 1. (Colour on-line) (a) Schematic representation of the fabrication process of the CuTCNQ nanowire device. The growth occurs laterally. (b) Top view SEM image of a Cu:TCNQ nanobridge device. Array indicates the direction of growth. The electrode from which growth starts is marked by star and direction of growth is marked by arrow. (c) Temperature dependent resistivity of the NW array.

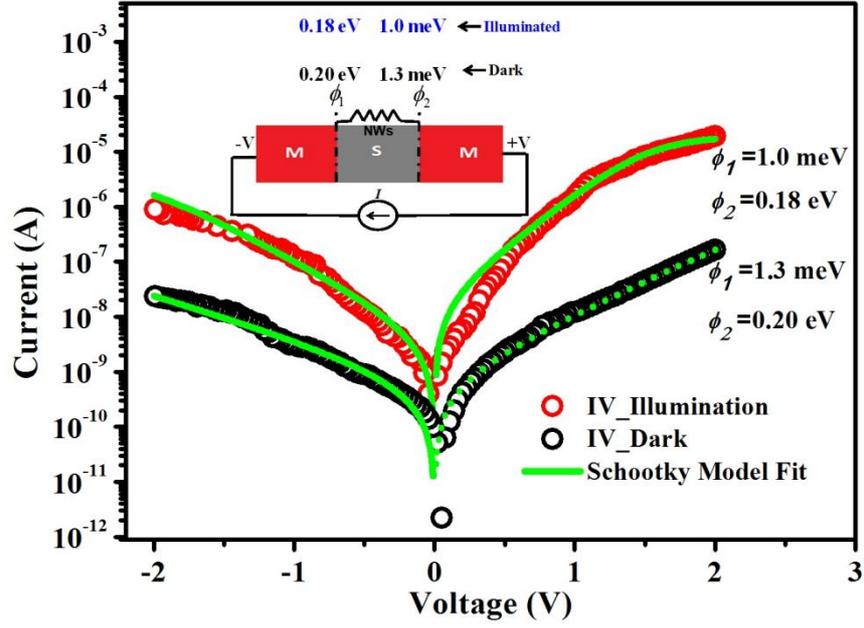

FIG. 2. (Colour online) *I-V* curves in dark (black circle) and under illumination (red circle) of Cu:TCNQ nanowire array. Wavelength of illumination = 405 nm with power density $P_{opt}= 2.0 \times 10^6$ W/m$^2$. Solid green curve is fit to the MSM device model. The barrier heights $\phi_1$ and $\phi_2$ obtained from the fit are $\phi_1 = 1.30 meV$, $\phi_2 = 0.20 eV$ in dark and $\phi_1 = 1.04 meV$, $\phi_2 = 0.18 eV$ under illumination. The inset shows the schematic of barriers at contacts.

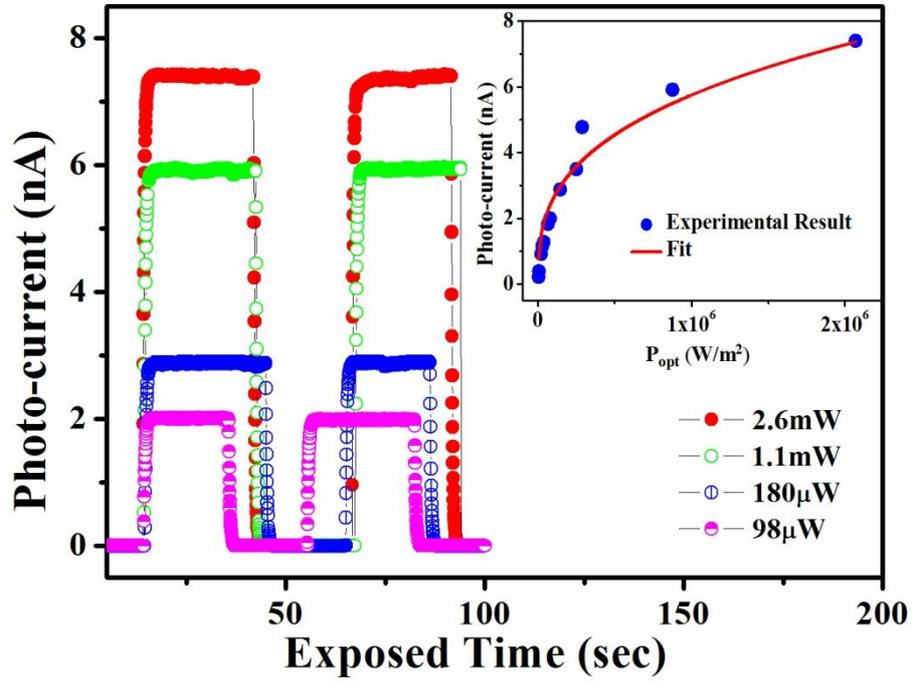

FIG. 3. Reversible zero bias photoresponse $I_{ph}$ ($V = 0$) of Cu:TCNQ nanobridge device under an illumination at wavelength 405nm with varying power density. Inset shows dependence of zero bias photocurrent on the incident optical power density $P_{opt}$.

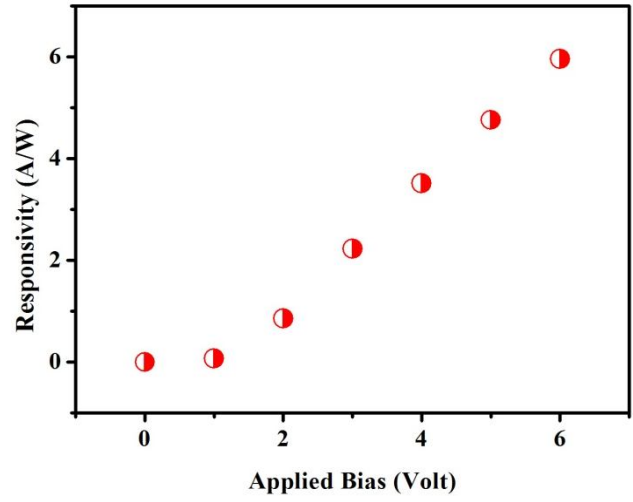

FIG. 4. Variation of responsivity of the array of NWs with applied bias. The number of wires in the specific nanobridge device is 80 and optical power density $P_{opt} = 6.6 \times 10^6 \, W/m^2$.